\documentclass[11pt,epsf]{article}
\newlength{\vshift}
\input epsf.tex
\newlength{\hshift}
\usepackage{amssymb}
\usepackage{amsmath}
\usepackage{amsthm}
\usepackage{amsopn}
\def\nn{\nonumber }
\def\la{\lambda}
\def\La{\Lambda}
\def\Si{\Sigma}
\def\th{\theta}

\def\de{\delta}
\def\De{\Delta}
\def\be{\beta}
\def\ga{\gamma}
\def\al{\alpha}
\def\si{\sigma}

\def\x{\hat x}
\def\p{\partial}

\def\lb{\lbrack}
\def\rb{\rbrack}
\def\beq{\begin{equation}}
\def\eeq{\end{equation}}
\def\bea{\begin{eqnarray}}
\def\eea{\end{eqnarray}}
\begin{document}

\vspace{4em}
\begin{center}

{\Large{\bf Massive Neutrino in Non-commutative Space-time}}

\vskip 3em

{{\bf  M. M. Ettefaghi \footnote{e-mail: ettefaghi@ph.iut.ac.ir } ,
M. Haghighat \footnote{e-mail: mansour@cc.iut.ac.ir }}}

\vskip 1em

Department of Physics Isfahan University of Technology, Isfahan,
84156-83111, Iran
 \end{center}
\begin{abstract}
We consider the noncommutative standard model based on $SU(3)\times
SU(2)\times U(1)$.  We study the gauge transformation of right
handed neutrino and its direct interaction with photon in the
noncommutative space-time.  We show that the massive Dirac
neutrinos, through the Higgs mechanism, can not accommodate this
extension of the standard model; while the massive Majorana
neutrinos are consistent with the gauge symmetry of the model.  The
electromagnetic properties and the dispersion relations for the
neutrino in the noncommutative standard model is examined.  We also
compare the results with the noncommutative standard model based on
$U(3)\times U(2)\times U(1) $.

\end{abstract}
%%%%%%%%%%%%%%%%%%%%%%%%%%%%%%%%%%%%%%%%%%%%%%%%%%%%%%%%%%%%%%%%%%%%%%%%%%%%%%%%%%%%%%%%%%%%%%%%%%%%%%%%%%%%%%%%%%%%%%%%%%%
\section{Introduction}
\label{A}By now there are convincing evidences which suggest
 neutrinos have finite masses such as the problem of the solar and atmospheric neutrinos and so on\cite{neutrino}.  Therefore
the extension of the standard model to include massive neutrinos is
natural.  But theoretically there are two different types of
neutrinos that are both neutral particles with different properties.
While the Dirac neutrinos like the other charged fermions are Dirac
particles, with the distinct antiparticles and the conservation of
lepton number, this is not the case for the Majorana neutrinos which
are the same as their antiparticles. Though the Majorana masses
violate the lepton number, the smallness of the neutrino masses
prevent us from confirming or excluding the Majorana property by
using the data of the processes sensitive to the lepton number
violation, whose typical example is the neutrino-less double
$\be$-decay. However, the violation of the total lepton number is
predicted by many extensions of the Standard Model (SM), therefore
the neutrinos can be Majorana as well. Consequently, the next
question is whether the existing neutrinos are Dirac or Majorana
particles or in other words the neutrino interactions within the SM
or beyond the SM lead us to distinguish them experimentally or even
suppress one of them theoretically.
  Indeed
the solar and atmospheric neutrino oscillations lead to the finite
masses for the neutrinos and their $\De m^2$ values and mixing
angles are known in such experiments with fair accuracy
\cite{fogli}. But, neutrino oscillation does not distinguish between
Dirac and Majorana neutrinos, because for example the equation for
the neutrino oscillations in vacuum is as follows
 \bea
i\frac{d}{dt}\begin{pmatrix}
  \nu_e \\
  \nu_\mu \\
  \nu_\tau
\end{pmatrix}\!\!\!\!&=&\!\!\!U\begin{pmatrix}
  \frac{m^2_1}{2E} & 0 & 0 \\
  0 & \frac{m^2_2}{2E} & 0 \\
  0 & 0 & \frac{m^2_3}{2E}
\end{pmatrix}U^\dag\begin{pmatrix}
  \nu_e \\
  \nu_\mu \\
  \nu_\tau
\end{pmatrix}\nn\\
&=&\!\!\!\frac{1}{2E}m_\nu m^\dag_\nu\begin{pmatrix}
  \nu_e \\
  \nu_\mu \\
  \nu_\tau
\end{pmatrix},\eea
where this expression holds for both pure Dirac and the neutrinos
appearing in the seesaw scenarios. In fact the most popular scenario
for the neutrino mass is the seesaw mechanism \cite{seesaw} in which
the small neutrino masses arise because of a large hierarchy between
a Dirac mass of the order of other fermions in the standard model
and a Majorana mass of the order of the Grand Unified mass scale or
at least several orders of magnitude larger than the standard model
scale.  There are however many attempts to find
a signal of new physics which can discriminate between Dirac and Majorana neutrinos \cite{dorm}. \\
On the other hand, the appearance of noncommutative (NC) field
theories in a definite limit of string theory \cite{sw} has provided
a strong motivation to investigate such theories. In recent years,
NC-field theories and their phenomenological aspects have been
explored by many authors \cite{nc}-\cite{nNCSM}.  In fact new
interactions
 in the NC space-time seem to be potentially important to particle physics and cosmology.
 In the canonical version of the NC space-time, the coordinates are operators and
satisfy the following relation
 \beq
\th^{\mu\nu}=-i\lb\x^\mu,\x^\nu\rb,
 \eeq
  where a hat indicates a
noncommutative coordinate and $\th^{\mu\nu}$ is a real, constant and
antisymmetric matrix.  To construct the noncommutative field theory,
according to the Weyl- Moyal correspondence, ordinary function can
be used instead of the corresponding noncommutative one by replacing
the ordinary product with the star product as follows
\begin{equation}
f*g(x)=f(x)exp(i/2\overleftarrow{\partial_{\mu}}\th^{\mu\nu}
\overrightarrow{\partial_{\nu}})g(x).
\end{equation}
Using this correspondence, however, there are two approaches to
construct the gauge theories in the noncommutative space. In the
first one the gauge group is restricted to $U(n)$ and the symmetry
group of the standard model is achieved by the reduction of
$U(3)\times U(2)\times U(1)$ to $SU(3)\times SU(2)\times U(1)$ by an
appropriate symmetry breaking \cite{nNCSM}. In the second approach,
the noncommutative gauge theory can be constructed for $SU(n)$ gauge
group via Seiberg-Witten map\cite{sm}.  In the both versions of the
NC field theories among many new interactions there is a direct
interaction between  a neutral particle and a photon.  For example
in the minimal extension of the standard model in the noncommutative
space there is neutrino-photon vertex which leads to neutrino-photon
interaction at the tree level \cite{hez}.  In this paper, our aim is
to study the effects of such interactions on the properties of
neutrino in vacuum.

In section 2 we briefly review the both versions of the
noncommutative standard model (NCSM).  In section 3 we study the
models for neutrino mass generation and show that the Dirac massive
neutrino through the Higgs
 mechanism is forbidden in the minimal version of NCSM (mNCSM) which is based on the gauge
group $SU(3)\times SU(2)\times U(1)$.  Meanwhile the massive
Majorana neutrino is consistent with both of the versions of NCSM.
We also show that to describe the electromagnetic current for the
neutrino, besides the usual form factors, new ones are needed. In
section 4 we consider the dispersion relation for the neutrino in
the NC space-time and show that the neutrino and its antiparticle
have generally different dispersion relation in the mNCSM. Finally
we summarize our results.

 %%%%%%%%%%%%%%%%%%%%%%%%%%%%%%%%%%%%%%%%%%%%%%%%%%%%%%%%%%%%%%%%%%%%%%%%%%%%%%%%%%%%%%%%%%%%%%%%%%%%%%%%%%%%%%%%%%%%%%%%%%%
\section{Non-commutative standard model}
There are two approaches to construct the standard model in the
noncommutative space. In the minimal extension of the standard model
in the noncommutative space, the gauge group is $SU(3)_c\times
SU(2)_L\times U(1)_Y$ in which the number of gauge fields, couplings
and particles are the same as the ordinary one \cite{sm}. Although
in this extension new interactions will appear due to the star
product and the SW map, the photon-neutrino vertex is absent.  We
denote the fermion content of the theory as
 \beq
  \widehat{L}=\left(
\begin{array}[]{c}
\widehat{\Psi}_{L_{\nu_l}}\\ \widehat{\Psi}_{L_l}
\end{array}
\right)\,\,\,\, , \,\,\,\,\left( \begin{array}[]{c}
\widehat{\Psi}_{L_u}\\ \widehat{\Psi}_{L_d}
\end{array} \right)
  \eeq
and
 \bea \widehat{R}\!\!\!\!&=&\!\!\!\!\widehat{\Psi}_{R_l}\,\,\,\,
, \,\,\,\,\widehat{\Psi}_{R_u}\,\,\,\, , \,\,\,\
\widehat{\Psi}_{R_d}\nn\\
\widehat{\nu}_R\!\!\!\!&=&\!\!\!\!\widehat{\Psi}_{R_{\nu_l}}\,\,\,,
 \eea
 where for three generations, $l$ stands for $e$, $\mu$ and $\tau$ meanwhile subscript $u$
refers to up-type quarks and subscript $d$ to down-type quarks. The
fields
 with hat are the non-commutative fields which can be written as a
 function of ordinary fields using appropriate SW-maps.
 Let us now consider an infinitesimal non-commutative local gauge
transformation of these fields as follows
 \beq
\delta \Psi=i\rho_\Psi(\Lambda) \star \Psi,
 \eeq
 where $\Lambda$ is a gauge parameter and $\rho_\Psi(\Lambda)$ is a
representation which is  carried by the matter or Higgs fields
according to the Table 1. In mNCSM, under the infinitesimal gauge
transformation $\widehat{L}$ and $\widehat{H}$ are transformed as
follows
 \bea\label{transformation}
\de\widehat{L}&=&i\widehat{\La}\star\widehat{L},\nn\\
\de\widehat{R}&=&i\widehat{\La}^\prime\star\widehat{R},\nn\\
\de\widehat{H}&=&i\widehat{\La}\star\widehat{H}-i\widehat{H}\star\widehat{\La}^\prime,
\eea
 in which the symbol $\rho_\Psi$ is omitted.  Meanwhile
 $\widehat{\nu}_R$ as a neutral-hyper-charged particle in the
 standard model transforms as
\bea\label{nutransformation}
\de\widehat{\nu}_R&=&i\widehat{\La}^\prime\star\widehat{\nu}_R-i\widehat{\nu}_R\star\widehat{\La}^\prime.
 \eea
 In fact in a non-commutative setting the non-commutative gauge
 boson ${B}_\mu$, compatible with the non-commutative gauge
 transformation, couples to a neutral matter field $\Psi^0$
 as follows
 \beq
\widehat{{\cal D}}_\mu \Psi^0=\p_\mu \Psi^0-i\lb{B}_\mu,\Psi^0\rb,
 \eeq
 with
 \beq
 \delta\Psi^0=i[\Lambda\stackrel{\star}{,}\Psi^0].
 \eeq
 In the second approach the gauge
group is $U_\star(3)\times U_\star(2)\times U_\star(1)$ which is
reduced to $SU(3)_c\times SU(2)_L\times U(1)_Y$ by reducing the two
extra $U(1)$ factors through the appropriate Higgs mechanism and
Higgs particles ({\it Higgsac})\cite{nNCSM}.
 In this approach the number of possible particles in each family (which are six:
left-handed lepton, right-handed charged lepton, left-handed quark,
right-handed up quark, right-handed down quark and Higgs boson)
 as well as their hyper-charges is naturally fixed.  Hence, in contrast to the mNCSM, the
right-handed neutrino could only be a sterile neutrino, i.e. a
singlet under all gauge groups. The gauge transformations for the
first generation of the leptonic part of the matter fields are as
the following:

{\bf 1) Right-handed leptons}. In this group there is only the
right-handed electron and its right-handed neutrino, which
transforms as
 \beq
 e_R (x)\ \rightarrow \ e_R (x)\, v^{-1} (x)\ ,
 \eeq
 and
 \beq\label{rnuGT}
 {\nu_e}_R (x)\ \rightarrow \ v(x) {\nu_e}_R (x)\, v^{-1} (x)\ ,
 \eeq
in which  $v$ is the elements of $U_\star(1)$.

{\bf 2) Left-handed leptons}. Here we have the the left-handed
electron and its neutrino, in a doublet:
 \beq\label{covlL}
  \Psi^l_L(x)=\left(\begin{array}{c}\nu (x)\\ e (x)\end{array}\right)_L\ .
   \eeq
Under the gauge transformations, the doublet transforms as
 \beq\label{lhGT}
\Psi^l_L (x)\ \to\ V(x)\, \Psi^l_L (x)\, v^{-1} (x)
 \eeq
where $V(x)$ is the elements $U_\star(2)$.

{\bf 3) Higgs doublet}
 \beq\label{Higgs}
 \Phi(x)=\left(\begin{array}{c}\Phi^+ (x)\\ \Phi^0 (x)\end{array}\right)\,
 \eeq
which transforms as
  \beq\label{PhiGT}
  \Phi(x)\ \to\ V(x)\,\Phi (x)\ ,
   \eeq
while the charge conjugated field of $\Phi$ transforms as:
\beq\label{PhiGT1}
  \Phi^c(x)\ \to\ V(x)\,\Phi^c (x) v^{-1}(x).
   \eeq

\begin{table}
\begin{center}
\begin{tabular}
 {|c|c|c|c|}\hline
    & $SU(3)_c$ & $SU(2)_L$ & $U(1)_Y$ \\ \hline
  $e_R$ & 1 & 1 & -2 \\ \hline
  $L_l=\begin{pmatrix}
    \nu_L \\
    e_L \
  \end{pmatrix}$ & 1 & 2 & -1 \\ \hline
  $u_R$ & 3 & 1 & $\frac{4}{3}$ \\ \hline
  $d_R$ & 3 & 1 & $\frac{-2}{3}$ \\ \hline
  $L_q=\begin{pmatrix}
    u_L \\
    d_L \
  \end{pmatrix}$ & 3 & 2 & $\frac{1}{3}$ \\ \hline
  $\Phi=\begin{pmatrix}
    \phi^+ \\
    \phi^0 \
  \end{pmatrix}$ & 1 & 2 & 1 \\ \hline$\nu_R$ & 1 & 1 & 0 \\ \hline
\end{tabular}
\end{center}
\caption{Matter and Higgs field content of the extended standard
model and their representation}
\end{table}

%%%%%%%%%%%%%%%%%%%%%%%%%%%%%%%%%%%%%%%%%%%%%%%%%%%%%%%%%%%%%%%%%%%%%%%%%%%%%%%%%%%%%%%%%%%%%%%%%%%%%%%%%%%%%%%%%%%%%%%%%%%
\section{Massive neutrino in non-commutative standard model}
There are several indications for nonzero neutrino masses of which
the most stringent ones come from the solar and atmospheric neutrino
experiments. It is obvious from the observed mass scale of the
neutrinos that they cannot be treated exactly the same way as the
other basic fermions.  There must be some reasons for them being
almost massless. In the standard model they are precisely massless
because there are no right-handed neutrino states.  In fact, as the
most straightforward way, one can construct Dirac mass terms for
neutrinos through the Higgs mechanism and Yukawa coupling.  Here one
encounters the smallness of the Yukawa coupling in comparison with
the other charged-fermions.  Indeed there is no natural explanation
for the smallness of the neutrino mass.
  If we consider the right
handed neutrinos, the Yukawa terms for neutrinos can be written in
the standard model similar to the other fermions as follows
  \beq\label{yukawa}
{Y_\nu}_{ij}{\bar{L}}_iH^c{\nu_R}_j+h.c.,
  \eeq
in which $H^c$ is the charged conjugate of the Higgs field. This
term  after the electroweak symmetry breaking leads to the usual
Dirac fermion mass i.e. ${{M_\nu}_D}_{ij}={Y_\nu}_{ij}\langle
H\rangle$.

   In the mNCSM, (\ref{yukawa}) leads to
 \beq\label{ncyukawa}
 {Y_\nu}_{ij}{\bar{\widehat{L}}}_i\star\widehat{H}^c\star{\widehat{\nu_R}}_j+h.c.,
  \eeq
  that under the gauge transformations which are given in Eqs.(\ref{transformation}-\ref{nutransformation})
 obviously violates the mNCSM gauge symmetry which is in contrast with the other
 fermions.  Therefore the massive Dirac neutrino through the Higgs
 mechanism is forbidden in the mNCSM.  Meanwhile, it should be noted
 that in the non-minimal version of NCSM \footnote{ Hereafter non-minimal version of NCSM is used for NCSM based on
 the gauge group $U_\star(3)\times U_\star(2)\times U_\star(1)$ and should not be confused with the
non-minimal versions of the NCSM based on the gauge group
$SU(3)_c\times SU(2)_L\times U(1)_Y$ which can be constructed due to
the freedom in the choice of the representation of the gauge group
in the pure gauge action\cite{sm}.  In this paper the minimal choice
for the gauge group $SU(3)_c\times SU(2)_L\times U(1)_Y$ is
introduced. }
  based on the  $U_\star(3)\times U_\star(2)\times U_\star(1)$ gauge
group, (\ref{ncyukawa}) is conserved, see (\ref{rnuGT}),
(\ref{lhGT}) and (\ref{PhiGT1}). However, as a prediction of almost
whole extensions of the standard model, the conservation of lepton
number can be broken.  Therefore neutrinos as neutral particles do
not hold any additive internal quantum numbers and it is possible to
introduce Majorana mass term for them as well. However, the seesaw
mechanism is not consistent with the mNCSM.
\subsection{Majorana mass for neutrino}
 In the both versions of the standard model in the noncommutative space neutrino can be a massive Majorana particle.  For the right
handed neutrino, the Majorana mass term can be written as
 \beq\label{Majorana}
 M_R\nu^T_RC^{-1}\nu_R+h.c.,
   \eeq
   where $C$ is the Dirac charge conjugation matrix. Since $\nu_R$  under the
 gauge group transforms as a singlet, (\ref{Majorana}) can be considered as a
bare mass term in the lagrangian or, alternatively, can be generated
by interactions with a singlet scaler field $\si$
 \beq \label{rhcupling}
f_\si\nu^T_RC^{-1}(\si+\langle\si\rangle)\nu_R,
 \eeq
 where $\langle\si\rangle$ is the vacuum expectation value of $\si$. Under
the mNCSM gauge group, (\ref{rhcupling}) is invariant if $\si$
transforms as
 \beq
  \de\widehat{\si}=
i\widehat{\La}^\prime\star\widehat\si-i\widehat\si\star\widehat{\La}^\prime,
 \eeq
while for the non-minimal version of NCSM it should be transformed
as
  \beq\label{siGT}
\si\ \rightarrow \ v(x) \si\, v^{-1} (x)\ .
 \eeq
%\item
To generate the Majorana mass for the left handed neutrinos, one can
use the non-renormalizable operator if the renormalizability of the
theory is abandoned \cite{weinberg}.  For this purpose one can write
 \beq\label{nrGT}
\frac{\la_{ij}}{M}(L_iH)^T(L_jH),   i,j=e,\mu,\tau,
 \eeq
 where $\la_{ij}$'s are the dimensionless couplings and $M$ is an appropriate
 cut-off.  However, to preserve the mNCSM symmetry in (\ref{nrGT}), the Higgs field should be transformed from the right hand side as a
singlet i.e. $\widehat{\La}^\prime=0$ in the transformation of the
Higgs field given in (\ref{transformation}).
  One can easily see that in the non-minimal NCSM, (\ref{nrGT}) under the symmetry transformations of this group is not conserved.\\
%\item
The Majorana mass term for the left handed neutrino violates lepton
number by 2 units and has the Weak isospin $I=1$ therefore it can be
generated due to coupling with the Higgs triplet  $\De$:
 \beq \label{De}
 f_\De L^TL\De+h.c..
  \eeq
  The non-zero vacuum expectation value of $\De$
then gives $m_L=f_\De\langle\De\rangle$. To conserve (\ref{De})
under the mNCSM gauge symmetry, $\De$ should be transformed as
 \beq
  \de\widehat{\De}=
-i\widehat{\La}\star\widehat\De+i\widehat\De\star\widehat{\La}.
 \eeq
Meanwhile (\ref{De}) is not conserve under the gauge transformations
of the non-minimal version of NCSM.
\subsection{neutrino dipole moment}
\indent Since neutrinos are neutral, they can couple with the
electromagnetic field in the ordinary space-time through loop
corrections. Therefore, the most general form for the
electromagnetic current between Dirac neutrinos, consistent with the
Lorentz covariance and the Ward identity, can be written as follows:
\bea \label{formfacors}
\langle\nu(p^\prime,\la^\prime)|J^{em}_\mu|\nu(p,\la)\rangle=\bar{\nu}&\!\!\!\!
(p^\prime,\la^\prime)\{\ga_\mu F_1(q^2)-\ga_\la\ga_5(g^\la_\mu
q^2-q^\la q_\mu)G_1(q^2)\nn\\&
+\si_{\mu\nu}q^\nu[F_2(q^2)-\ga_5G_2(q^2)]\}\nu(p,\la),\eea
 where
$q=p^\prime-p$ and $F_1$, $G_1$, $F_2$ and $G_2$ are electric
charge, anapole (axial charge), magnetic and electric form factors,
respectively. All of the form factors are nonzero for a massive
Dirac neutrino if the CP violation is included.  In fact even if the
leptonic sector of the weak interaction conserve CP in the quark
sector, CP is not conserved and at least at high order loops
involving virtual quarks it will be induced.  In the case of
massless neutrino, the matrix elements of electromagnetic current
can be expressed in terms of only one form factor as \cite{rosado}
\beq\langle\nu(p^\prime,\la^\prime)|J^{em}_\mu|\nu(p,\la)\rangle=
F(q^2)\bar{\nu}(p^\prime,\la^\prime)\ga_\mu(1+\ga_5)\nu(p,\la). \eeq
 Meanwhile, for Majorana neutrinos which are the
same as their antiparticles the only nonzero form factor is the
anapole.  However the transition matrix elements relevant to
$\nu_i\rightarrow\nu_j$ would still generally contain four form
factors as given in (\ref{formfacors})
 for both the Dirac and Majorana neutrinos. \\
\indent In the non-commutative space time the parameter
$\theta^{\mu\nu}$ carries Lorentz indices but it does not mean that
it is a Lorentz quantity under a general Lorentz transformation. In
fact there are two distinct types of Lorentz transformation which
are observer and particle Lorentz transformations \cite{lorentz
trans}. Nonetheless, $\theta^{\mu\nu}$ is only a Lorentz tensor
under the observer Lorentz transformation while it is a constant
under the particle Lorentz transformation. Therefore the Lagrangian
in the NC space is fully covariant under the observer Lorentz
transformations.  This means that the observer Lorentz symmetry is
an invariance of the model, but the particle Lorentz group is
broken. Hereafter Lorentz quantity means the Lorentz quantity under
the observer Lorentz transformation.  However, there is a new
Lorentz quantity in the NC-space beside the usual ones i.e.
$\th_{\mu\nu}$. Therefore for the Dirac neutrinos, the effective
$\nu\nu\ga$ vertex can be generally expanded in terms of the Lorentz
vectors as follows:
 \bea\label{ncformfac}
\langle\nu(p^\prime,\la^\prime,\th)|J^{em}_\mu|\nu(p,\la,\th)\rangle
= \bar{\hat{\nu}}(p^\prime,\la^\prime)&&\!\!\!\!\!\!\!\!\!\!\!
\{\ga_\mu F_1-\ga_\la\ga_5(g^\la_\mu q^2-q^\la
q_\mu)G_1+\si_{\mu\nu}q^\nu[F_2-\ga_5G_2]\nn\\
&&+\th_{\mu\nu}\ga^\nu(F_3+\ga_5G_3)+\th_{\mu\nu}q^\nu(F_4+\ga_5G_4)\nn\\
&&+F_5(\th_{\mu\nu}\si^{\nu\rho}-\th^{\rho\nu}\si_{\nu\mu})q_\rho\nn\\
&&+G_5(\th_{\mu\nu}\si^{\nu\rho}-\th^{\rho\nu}\si_{\nu\mu})\ga_5q_\rho\}\hat{\nu}(p,\la)\,,
\eea where the form factors $F_i$'s and $G_i$'s depend on the
Lorentz invariant quantities such as $q^2$ and
$q^\mu\th_{\mu\nu}p^\nu$.  Here one has:
 \begin{enumerate}
 \item  The Ward identity leads to $F_3=G_3=0$.
 \item  Up to the first order of $\th_{\mu\nu}$ only $F_i$'s and
 $G_i$'s with $i=1,2$ have $\th$-dependent parts.
 \item  Since $\nu=\nu_L+\nu_R$ the current in (\ref{ncformfac}) is
 left-left for $F_1$ and $G_1$ while it is left-right for $F_i$'s and
 $G_i$' with $i=2,4,5$.  Therefore the corresponding effective
 Lagrangian contains interactions which involve the neutrino
 bilinear $\bar\nu_L \Gamma_\mu \nu_L$ and $\bar\nu_L \Gamma^\prime_\mu
 \nu_R$ multiplied by a functional of the hypercharge field.
 \end{enumerate}
In the mNCSM those terms which are proportional to $\bar\nu_L
\Gamma^\prime_\mu \nu_R$ obviously violate the mNCSM gauge symmetry,
see Eqs.(\ref{transformation}-\ref{nutransformation}). Therefore, in
contrast with the non-minimal version of NCSM, the current in
(\ref{ncformfac}) can be written only in terms of $F_1$ and $G_1$.
  In the mNCSM, the fermion fields depend on the parameter of
  the non-commutativity of space and time as follows
   \bea \label{epsi}
{\widehat{\Psi}}=\Psi^0+ \Psi^1 +\Psi^2+\dots,
 \eea
 where $ \Psi^0$ is the commutative fermion
field and the superscript $i=1,2,...$ stands for the order of
$\th_{\mu\nu}$ in the expansion. For instance
 \beq
\Psi^1=-\frac{1}{2}\th^{\mu\nu}A^0_\mu\p_\nu\Psi^0+\frac{i}{4}\th^{\mu\nu}A^0_\mu
A^0_\nu\Psi^0,\label{psi1}
 \eeq
 where $A^0$ is the ordinary gauge field \cite{ncg}.  Under charge conjugation
 operation ($\cal C$), a free fermion field in the noncommutative space-time transforms as
 \bea \label{epsi-c}
 {\cal C} {\widehat{\Psi}}{\cal C}^{-1}=\ga_0C{\widehat{\Psi}}^*.
 \eea
 Since for the Majorana neutrino, ${\widehat{\Psi}}^c={\widehat{\Psi}}$, we should have
 \beq \label{ncth}
 {\cal C}^{-1}\th_{\mu\nu}{\cal C}=-\th_{\mu\nu},
  \eeq
  which is consistent with the transformation property of
  $\th_{\mu\nu}$ under ${\cal C}$ for the both versions of
  NCSM which are given in \cite{cpt} and \cite{gut}.  One should note that
    the operator $C$ in (\ref{epsi-c}) contains the $\th$ transformation in addition to the usual
  $i\ga_2\ga_0$.
  Therefore, in the current (\ref{ncformfac}) for the Majorana
  neutrinos, up to the first order of
  $\th_{\mu\nu}$, besides the anapole, the $\th$-dependent part of $F_1$, $F_2$, $G_2$
  and $F_5$ and $G_5$ are nonzero.  Here one should note that the
  all form factors would generally survive in the transition
 $\nu_i\rightarrow\nu_j$ for the Majorana neutrinos which are consistent with the both versions of NCSM.

 In the ordinary space neutral particles can only interact with
 photon through the loop corrections which is in contrast with the non-commutative space
  where they can minimally couple to each other, even at the tree level, in the adjoint representation of $U_\star(1)$ gauge
group. For instance in the mNCSM only neutral particles with zero
 hypercharge can couple directly to the photon, therefore the left handed neutrinos cannot couple to the hyperphoton.  In fact if we consider
 the right handed neutrino as a particle in the model, its interaction with the photon can be obtained from the following Lagrangian
\cite{hez}:
 \bea\label{hez}
&{\cal{L}}_{\nu_R}&=i{\bar{\nu}}_R\partial\!\!\!/\nu_R +
ie{\theta}^{\mu\nu}[\partial_\mu{\bar{\nu}}_RB_\nu\gamma^\rho({\partial}_\rho\nu_R)
\nonumber\\& &
-\partial_\rho{\bar{\nu}}_RB_\nu\gamma^\rho({\partial}_\mu\nu_R) +
{\bar{\nu}}_R( \partial_\mu B_\rho
)\gamma^\rho({\partial}_\nu\nu_R)]\label{Lagden}, \eea where $B$ in
terms of the photon and the $Z$-gauge boson fields is
\begin{equation}
B= \cos\theta_W A-\sin\theta_W Z,
 \end{equation}
 Which leads to the $\nu\nu\ga$ vertex as follows
\beq
\Gamma^{\mu}=i{\frac{1}{2}}cos\th_W(1+\gamma_5)(\th^{\mu\nu}k_\nu
q\!\!\!/ + \th^{\rho\mu}q_\rho k\!\!\!/ + \th^{\nu\rho}k_\nu q_\rho
\gamma_\mu).
 \eeq
Meanwhile in the non-minimal version of NCSM the left handed
neutrinos, as well as the right handed neutrinos, can be coupled to
photon through the following interaction \cite{nNCSM}:
 \beq
\label{enmncsm}{\cal
L}_{\nu-\gamma}=-ie\bar{\nu}\gamma^{\mu}[\nu,A_{\mu}]_{\star}=-e\
\bar{\nu}\gamma^{\mu}\left(\theta_{\alpha\beta}\partial_{\alpha}A_\mu
\partial_{\beta}\nu\right) +{\cal O}(\theta^2)\ .
 \eeq
Therefore in contrary to the commutative space where the form
factors $F_1$ and $G_1$ vanish for the arbitrary neutrino mass in
the 't Hooft-Feynman gauge \cite{dvornikov}, these form factors
which depend on the lorentz invariant quantity $p.\th.q$, are
nonzero at the leading order even for the massless neutrinos.
Moreover, comparing $G_1$ induced by the noncommutative space with
the electroweak anapole defined through the $\nu_l l$ scattering at
the one loop level \cite{rosado}, we find
$\La_{NC}=\frac{1}{\sqrt{\th^2}}\geq 10 TeV$.

%%%%%%%%%%%%%%%%%%%%%%%%%%%%%%%%%%%%%%%%%%%%%%%%%%%%%%%%%%%%%%%%%%%%%%%%%%%%%%%%%%%%%%%%%%%%%%%%%%%%%%%%%%%%%%%%%%%%%%%%%%%
\section{Dispersion relation of neutrino in vacuum}
In the previous section we showed that neutrino properties in the
noncommutative space are different from the ordinary space.  In this
section we show how the new interaction of neutrino in the
noncommutative space-time can affect its dispersion relation in
comparison with the usual space.  To this end we explore the pole of
the neutrino-propagator in vacuum of the NC-space-time. In the
ordinary space-time the Lagrangian of a free massless neutrino in
 the momentum space and in the vacuum can be written as follows:
  \beq {\cal
L}_0=\bar{\psi}_L(p)p\!\!\!/\,\psi_L(p).
 \eeq
  By including the interactions, the self-energy of the neutrino can be easily obtained as
   \beq
\label{usual} {\cal L}=\bar{\psi}_L(p)(p\!\!\!/\,-\Si)\psi_L(p),
 \eeq
  where $\Si$, in the vacuum, is generally given by:
 \beq \Si=ap\!\!\!/,
  \eeq
  in which  $a$ is a lorentz scalar which depends on the lorentz invariant quantities. However,
  in addition to the momentum of the
neutrino, in terms of the parameter of the noncommutativity, new
4-vectors such as $\th_{\mu\nu}p^\nu$  and
$\th_{\mu\nu}\th^{\nu\rho}p_\rho$ can be constructed in the
noncommutative space-time. Therefore, the most general form of the
$\Si$ up to the second order of $\th$ is as follows:
 \beq\label{si}
\Si=a p\!\!\!/+b p.\th\!\!\!/+c p.\th.\th\!\!\!/\,\,,
 \eeq
  where $a$, $b$ and $c$ are Lorentz scalars and depend on the corresponding Lorentz invariant quantities in the NC-space-time and
$p.\th\!\!\!/$ and $p.\th.\th\!\!\!/$ stand for
$p^\mu\th_{\mu\nu}\ga^\mu$ and
$p^\mu\th_{\mu\nu}\th^{\nu\rho}\ga_\rho$, respectively. The
Lagrangian (\ref{usual}) can be rewritten in the form:
 \beq {\cal
L}=\bar{\psi}_L(p)\frac{V\!\!\!\!/\,}{V^2}\psi_L(p),
 \eeq
  where
   \beq
V_\mu=(1-a)p_\mu-b p.\th_\mu-c p.\th.\th_\mu\,\,\,.
 \eeq
  The dispersion relation can be easily obtained by solving the equation $V^2=0$,
i.e.
 \beq \label{dis}(1-a)^2p^2-(2c(1-a)+b^2)p.\th.\th.p=0 .
  \eeq
   It is clear from (\ref{dis}) that the solutions for the
positive and negative energy and therefore the dispersion relations
for the neutrino and antineutrino are different only if
$\th^{i0}\neq 0$. In fact if we construct a consistent field theory
on the noncommutative space-time, the neutrino and its antiparticle
can be distinguished in the vacuum by the nonzero value of
$\th^{i0}$
which is in contrast with the ordinary space-time.\\
\begin{figure}\label{diag}\centerline{\epsfysize=1.9
in\epsfxsize=3.5 in\epsffile{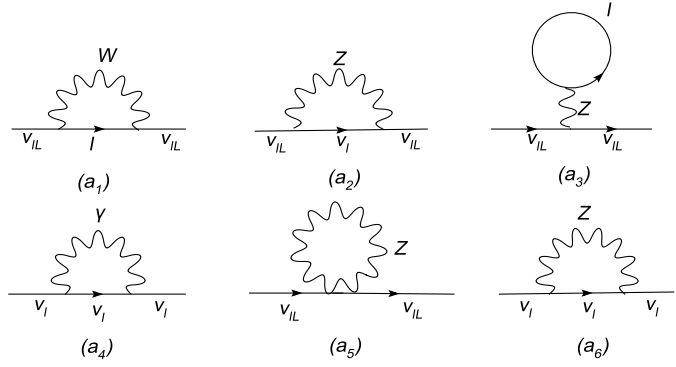}} \caption{1-loop diagrams
for neutrino self-energy in noncommutative space}
\end{figure}
Now we calculate the explicit form of $a$, $b$ and $c$ for the both
versions of the NCSM. The relevant diagrams for the neutrino self
energy in the NC-space are shown in figure (1). By expanding the
propagator of the W-gauge boson as a power series in $1/M_W^2$:
 \beq
D_{\al\be}(q^2)=\frac{-g_{\al\be}+\frac{q_\al
q_\be}{M_W^2}}{q^2-M^2_W}=\frac{g_{\al\be}}{M_W^2}+\frac{q^2g_{\al\be}-q_\al
q_\be}{M_W^4}+\ldots\,,
 \eeq
  and using the dimensional regularization which leads to:
   \beq \int d^dq
(q^2)^\nu=0\,\,\,\,; \text{ for any $\nu$ },
 \eeq
  one can easily see that the contributions from the diagrams $a_1$, $a_2$, $a_3$, $a_5$ and $a_6$
 are zero.  The diagram  $a_4$ which is absent in the ordinary space can be obtained in the mNCSM by using
 the neutrino-hypercharge vertex given in (\ref{hez}) as
 \beq\label{a4}
 \Si=
fp^2(2p^2p.\th.\th.\ga+(p.\th.\th.p-\frac 1 2
\th^{\mu\nu}\th_{\mu\nu}p^2)p\!\!\!/)R,
 \eeq
  where
$R=\frac{1+\ga_5}{2}$ and
 \beq\label{T}
f=\frac{ie^2}{12(4\pi)^2}(1-\frac{1}{\varepsilon}-ln(\frac{\mu^2}{p^2})-\ga-ln4-\psi(\frac
5 2)).
 \eeq
 Comparing (\ref{a4}) with (\ref{si}) leads to $b=0$ and
\beq\label{a}
 a=
fp^2(p.\th.\th.p-\frac 1 2 \th^{\mu\nu}\th_{\mu\nu}p^2)R,
 \eeq
 and
 \beq\label{c}
 c=
2fp^4R.
 \eeq
Meanwhile the diagram  $a_4$ in the non-minimal version of NCSM can
be easily calculated by considering the vertex given in
(\ref{enmncsm}) which leads to:
 \beq\label{enma4}
  \Si=-f p.\th.\th.p
p^2p\!\!\!/,
 \eeq
  where $b=c=0$ and $a=-f p.\th.\th.p
p^2$. Since $a$ can be absorbed in the field redefinition, the
effects of the noncommutativity on the dispersion relation can be
detected only in the mNCSM.

%%%%%%%%%%%%%%%%%%%%%%%%%%%%%%%%%%%%%%%%%%%%%%%%%%%%%%%%%%%%%%%%%%%%%%%%%%%%%%%%%%%%%%%%%%%%%%%%%%%%%%%%%%%%%%%%%%%%%%%%
\section{Conclusion}
In this paper we have studied the effects of the photon neutrino
interaction in NCSM on the neutrino properties in the vacuum.  In
the minimal version of the NCSM we have found:
\begin{enumerate}
 \item The Yukawa terms for neutrinos in the mNCSM, see (\ref{ncyukawa}), violates the mNCSM gauge
 symmetry, therefore the massive Dirac neutrino through the Higgs
 mechanism is forbidden and the seesaw mechanism is not consistent with the
 mNCSM.
 \item The mass term for Majorana neutrino can be defined directly,
 see (\ref{Majorana}), generated
  by interactions with a singlet scaler field $\si$, see
 (\ref{rhcupling}), through coupling with the Higgs triplet  $\De$, see
 (\ref{De}), or even through the non-renormalizable operator, see
 (\ref{nrGT}).
 \item The effective $\nu\nu\ga$ vertex, in the NCSM, for Dirac neutrinos can be
 generally expanded  in terms of 8 form factors in contrast with 4
 in the ordinary space, see (\ref{ncformfac}).  Those terms in the current which correspond to the interactions  proportional to $\bar\nu_L
\Gamma^\prime_\mu \nu_R$ in the effective
 Lagrangian violate the mNCSM gauge symmetry.  Therefore, the current
 in the mNCSM can be written only in terms of two form factors
 $F_1$ and $G_1$, see (\ref{ncformfac}) and (\ref{transformation}-\ref{nutransformation}).
 \item For the Majorana neutrino to be its antiparticle, the
 parameter of noncommutativity under charge conjugation
 operation $\cal C$ should transform as ${\cal C}^{-1}\th_{\mu\nu}{\cal
 C}=-\th_{\mu\nu}$, see (\ref{ncth}).  In fact in the NC-space
  the charge conjugation operator contains the $\th$ transformation in addition to the usual
  $i\ga_2\ga_0$. Therefore the electromagnetic current for the Majorana
  neutrino should be described in terms of 5 form factors $F_1$, $F_2$,
  $G_2$, $F_5$ and $G_5$ in contrast with the one, the anapole, in the commutative
  space.
 \item In contrary to the commutative space in the both version of
 the NCSM, $F_1$ and $G_1$ are nonzero at the leading order even for the massless neutrinos, and
  depend on the lorentz invariant quantity $p.\th.q$, see (\ref{hez}) and (\ref{enmncsm}), which lead to the upper bound $\La_{NC}=\frac{1}{\sqrt{\th^2}}\geq 10 TeV$.
 \item The dispersion relations
for the neutrino and antineutrino in the NC space-time are generally
different only if $\th^{i0}\neq 0$, see (\ref{dis}).  The self
energy for the neutrino can be expanded in terms of the new
$\th$-dependent four vectors up to the second order of
$\th_{\mu\nu}$ as $\Si=a p\!\!\!/+b p.\th\!\!\!/+c
p.\th.\th\!\!\!/\,\,$ where for the mNCSM $c\neq 0$, $a\neq 0$ and
$b=0$, see (\ref{si}) and (\ref{a4}). In other words, the neutrino
and its antiparticle have different dispersion relations in the
mNCSM.
 \end{enumerate}
 Meanwhile for the non-minimal version of NCSM we found:
 \begin{enumerate}
 \item The Yukawa term for neutrino in the non-minimal version of NCSM, (\ref{ncyukawa}), is consistent with the gauge
 symmetry given in (\ref{rnuGT}), (\ref{lhGT}) and (\ref{PhiGT1}), therefore the massive Dirac neutrino through the Higgs
 mechanism is allowed.
 \item In the non-minimal version of NCSM the mass term for Majorana neutrino can be defined
 directly, see
 (\ref{Majorana}), or  generated
  by interactions with a singlet scaler field $\si$, see
 (\ref{rhcupling}).
 \item Eight form factors are needed to describe the current for the
 Dirac neutrinos in the non-minimal version of NCSM and there is not any restriction by the gauge
 symmetry.
 \item The dispersion relations
for the neutrino and its antiparticle in the non-minimal version of
NCSM are the same because $c=b= 0$ and $a\neq 0$, see (\ref{si}) and
(\ref{enma4}). One should note that at our disposal we expanded the
non-local lagrangian of the NCSM based on the gauge group
$U_\star(3)\times U_\star(2)\times U_\star(1)$, to obtain the
result.  However, the expansion of the lagrangian in terms of
$\theta^{\mu\nu}$ destroys the non-locality of the model and it
should be shown that if the non-locality of the model has any
contribution on the obtained result or not.

    \end{enumerate}
\section*{Acknowledgements}
The financial support of Isfahan University of Technology research
council is acknowledged.

%%%%%%%%%%%%%%%%%%%%%%%%%%%%%%%%%%%%%%%%%%%%%%%%%%%%%%%%%%%%%%%%%%%%%%%%%%%%%%%%%%

\end{document}